\documentclass[
a4paper,
twocolumn,
amsmath,
aps,
prl, reprint,showpacs,
]{revtex4-1}

\usepackage[english]{babel}
\usepackage[utf8]{inputenc}
\usepackage{graphicx}
\usepackage{epstopdf}
\usepackage[colorinlistoftodos]{todonotes}

\date{\today}

\begin{document}

\title{Coherent buildup of high harmonic radiation: the classical perspective}

\date{\today}

\author{S. A. Berman$^{1,2}$, J. Dubois$^2$, C.\ Chandre$^2$, M.\ Perin$^1$, T.\ Uzer$^1$}
\affiliation{$^1$School of Physics, Georgia Institute of Technology, Atlanta, Georgia 30332-0430, USA}
\affiliation{$^2$Aix Marseille Univ, CNRS, Centrale Marseille, I2M, Marseille, France}


\begin{abstract}
We present a classical model for high harmonic generation during the propagation of an intense laser pulse through an atomic gas. Numerical simulations of the model show excellent quantitative agreement with the corresponding quantum model for the blueshift and intensity reduction of the propagating laser pulse over experimentally realistic propagation distances.  We observe a significant extension of the high-harmonic cutoff due to propagation effects. Phase-space analysis of our classical model uncovers the mechanism behind this extension.
\end{abstract}

\maketitle

High harmonic generation (HHG) is the production of coherent high-frequency radiation observed during the ionization of gases by intense laser pulses.
The high-frequency part of the spectrum typically consists of a plateau region, where the harmonics are produced with comparable intensity, followed by a cutoff region, which is often harnessed to generate attosecond pulses.
To increase the flux of the highest harmonics for more intense attosecond pulses, experiments may be performed under conditions of increased driving laser intensity \cite{Haes14}, gas length \cite{Kaza03}, or gas density \cite{Cons99,Popm12,Popm15}.
Under such conditions, the driving laser field undergoes tremendous reshaping while propagating through the gas due to the radiation emitted by the ionizing atoms, leading for example to a blueshift and intensity reduction \cite{Rae94,Geis99,Gaar06} throughout propagation.
In this case, the high harmonic spectrum measured after propagation crucially depends on which high harmonic frequencies were produced at sufficient intensity all along the gas with just the right phase such that the radiation produced by the many atoms making up the gas adds up coherently, a collective effect referred to as phase-matching \cite{Gaar08,Popm10}.
Thus, the self-consistent interaction between the ionizing atoms and the laser field plays a decisive role in shaping the high harmonic spectrum \cite{Cons99,Popm10,Popm12}.

Ideally, a theoretical or numerical treatment of HHG must bridge the gap between the microscopic response of the atoms to the electromagnetic field and the macroscopic propagation of the field through a gas of billions of atoms.
The most rigorous calculation would require the self-consistent solution of Maxwell's equations in three dimensions coupled to time-dependent Schr\"odinger equations (TDSEs) for the atoms \cite{Lori12,Gaar11}.
Even today, the computational cost associated with this approach can be prohibitive, precluding the simulation of experimentally relevant sample lengths on the order of $\mathrm{mm}$s.
Further, solutions of the TDSE provide limited intuition into the electron dynamics behind the single-atom response to the laser field.
Alternatively, one can simplify the description of the atomic response, splitting it into a low-frequency part dominated by ionization \cite{Geis99} and a high-frequency part \cite{Brab00,Milo02} comprised of the radiation emitted during repeated encounters between the ionized electrons with their parent ions \cite{Kula93,Cork93}.
The latter may be computed efficiently \cite{Gaar08} using a semiclassical approach \cite{Lewe94,Sali01} under the assumption that the ionic core potential has a negligible effect on the ionized electron dynamics.
This framework allows the simulation of experimental gas lengths, and the semiclassical description of the atomic response in terms of quantum orbits \cite{Lewe94,Sali01} facilitates the development of control strategies based on the trajectories of electrons after ionization \cite{Chip09,Haes14,Briz13}.
However, these simplifications are inappropriate for the description of the harmonics near and below the atomic ionization threshold $I_p$ \cite{Yost09,Xion14,Xion17}, which in certain situations can strongly influence the yield of higher harmonics \cite{Briz13,Gaar05}.
Additionally, they leave out key elements of HHG in elliptically and circularly polarized fields \cite{Shaf12,Aban17}.
Therefore, a theoretical formulation is needed which simultaneously accounts for the full complexity of the self-consistent atom-field interaction, includes the influence of the core potential on the ionized electrons, and allows for the understanding of the electron dynamics in phase space as the pulse propagates through the gas.
Here, we propose a purely classical model which meets these requirements, and we demonstrate its validity and utility by comparing its behavior with a quantum model.
In particular, we use it to identify the mechanism behind an intriguing phenomenon --the extension of the cutoff-- observed in quantum simulations.

\begin{figure}
\includegraphics[width=0.45\textwidth]{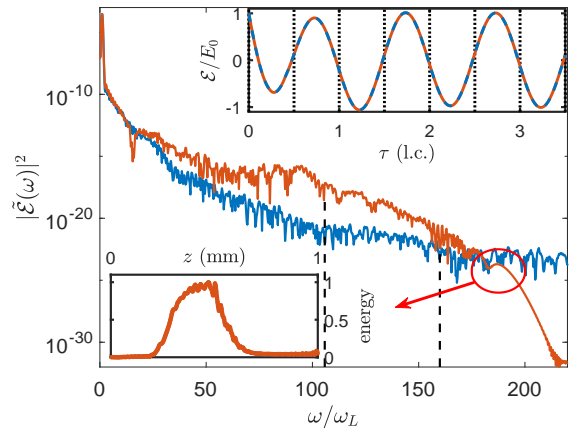}
\caption{Filtered electric field spectra of a $3.5$ laser cycle (l.c.) pulse at $z = 0.37\,\, \mathrm{mm}$ of a singly ionized $1\,\,\mathrm{mm}$ gas with density $\rho = 5\times10^{17}\,\,\mathrm{cm}^{-3}$. The incident field is $\mathcal{E}(\tau,z=0) = E_0 \cos(\omega_L \tau)$, with the field amplitude $E_0$ and frequency $\omega_L$ corresponding to an intensity $I=3.5\times10^{14}\,\,\mathrm{W}\cdot\mathrm{cm}^{-2}$ and wavelength $\lambda=1.2\,\,\mu\mathrm{m}$. The softening parameter $a^2=2$ is chosen to correspond to an ionization potential $I_p=0.5\,\,\mathrm{a.u.}$ The electron wavepacket is initialized at the quiver radius. For the blue curve the dipole velocity is computed classically, while for the orange curve it is computed quantum-mechanically. The dashed black lines correspond to $2\,U_p + I_p$ and $3.17\,U_p + I_p$. Upper inset: The time-dependent electric fields from each model at $z = 1\,\,\mathrm{mm}$. The black dotted lines indicate the locations of the extrema of the initial field. Lower inset: Energy (in arbitrary units) in the electric field frequency band between $175\omega_L$ and $200 \omega_L$, as a function of propagation distance $z$. 
}\label{fig:spectra}
\end{figure}

Our model describes the coupled evolution of the time-dependent electric field $\mathcal{E}(\tau,z)$ and the response of the atoms of the gas to the field throughout the laser pulse propagation.
The evolution parameter in our model is $z$, the position along the laser propagation direction.
For the atoms located at $z$, $\tau$ is the time relative to the arrival of the laser pulse to their position, i.e. $\tau = t-z/c$.
We may specify an arbitrary initial state for the atoms at $\tau=0$, which is uniform throughout the gas, and we may calculate the response of the atoms to the field in a classical or quantum manner.
In what follows, we consider a linearly-polarized laser pulse interacting with single-active-electron (SAE) atoms.
As an example, we present simulations of the model for a laser pulse with initial condition $\mathcal{E}(\tau,z=0) = E_0 \cos(\omega_L \tau)$ propagating through $1\,\,\mathrm{mm}$ of a gas with density $\rho=5\times10^{17}\,\mathrm{cm}^{-3}$.
The atomic initial condition, i.e.\ at $\tau=0$, is that of Refs.~\cite{Prot96,Sand99} - a fully ionized state with the electron described by a Gaussian wavepacket at rest at a distance of the quiver radius $E_0/\omega_L^2$ from the ion, expressed in atomic units which are used unless stated otherwise.
In Fig.~\ref{fig:spectra}, we compare the power spectra of the electric field at $z=0.37\,\,\mathrm{mm}$, using a classical description of the atoms on the one hand and a quantum-mechanical one on the other.
The spectra coincide at low frequencies, especially near the laser fundamental $\omega_L$, which is the frequency range where the dominant propagation effects on the electric field --the blueshift and intensity reduction-- are encoded.
In the upper inset of Fig.~\ref{fig:spectra}, these effects are clearly seen on the time-dependent electric field after propagation through $1\,\,\mathrm{mm}$ of gas: the blueshift is evident from the shift of the field extrema to the left of their initial positions (indicated by dotted lines), while the intensity reduction is seen on the leading edge of the pulse, where the absolute value of the field amplitude relative to the incident amplitude is less than one.
These effects are captured equally well by the classical and quantum atomic models: their respective time-dependent electric fields are indistinguishable in the upper inset of Fig.~\ref{fig:spectra} and differ by less than $10^{-2}E_0$ for all considered times $\tau$ and propagation distances $z$.
On the other hand, the classical description does not capture the high-harmonic plateau and cutoff radiation, which is present in the quantum model.
There, we also observe a significant extension of the high-harmonic cutoff past the $3.17U_p + I_p$ cutoff law, where $U_p=E_0^2/4\omega_L^2$ is the ponderomotive energy.
This is unexpected, given that the incident laser field is monochromatic and each atom only has a single active electron.
Radiation at these anomalously high frequencies only begins to emerge clearly about $0.2\,\,\mathrm{mm}$ into the gas, shown in the lower inset of Fig.~\ref{fig:spectra}, indicating that it is truly a propagation effect.
In the following, we will show that the purely classical model actually allows us to understand the mechanism of this anomalous high-harmonic radiation, despite its failing to capture the high-harmonic part of the spectrum on a quantitative level.
%

To begin, we consider the physics behind the classical model.
We derived the model from first principles, starting from Maxwell's equations and the Lorentz force law for classical charged particles.
The SAE atoms are assumed to have a static ionic core, and the electrons are assumed to be nonrelativistic and moving only in the direction transverse to the laser propagation direction $z$.
Meanwhile, the electric field is assumed to lie in the polarization plane, with no longitudinal component, and its only spatial dependence is assumed to be the propagation coordinate $z$.
Thus, our model neglects three dimensional effects, in particular the focusing of the laser beam and thus the Gouy phase shift.
If desired, a $z$-dependent phase and intensity may be imposed externally to partially account for these effects \cite{Kim02}, though we choose not to do this here in order to emphasize the self-consistent interaction between the radiation and the particles.
Lastly, we assume that backward-propagating waves may be neglected, i.e.\ the field propagates solely in the positive $z$ direction.
Under these assumptions, the evolution equation for the electric field may be written in a frame moving at the speed of light $c$ with the incident laser pulse as 
\begin{equation}
\frac{\partial \mathcal{E}}{\partial z} = \frac{2\pi \rho}{c} \langle v(\tau,z) \rangle,
\end{equation}
where $\rho$ is the atomic density and $\langle v(\tau,z) \rangle$ is the ensemble-averaged dipole velocity at time $\tau$ of the atoms located at $z$ driven by the field $\mathcal{E}(\tau,z)$.
For simplicity, we consider $\mathcal{E}$ to be linearly polarized along the $x$-direction as it is at $z=0$ and take one-dimensional models for the atoms, but the two dimensional generalization is straightforward.
For the classical model, $\langle v(\tau,z) \rangle = \int v f(x,v,\tau; z) \mathrm{d}x\mathrm{d}v $, where $f$ is the probability distribution function to find an electron with position $x$ relative to the ionic core and velocity $v$.
At every $z$, $f$ satisfies the Liouville equation
\begin{equation}\label{eq:liouville}
\frac{\partial f}{\partial \tau} = - v \frac{\partial f}{\partial x}+ \left( \frac{\partial V}{\partial x} + \mathcal{E}(\tau,z) \right) \frac{\partial f}{\partial v},
\end{equation}
corresponding to the single-atom Hamiltonian
\begin{equation}\label{eq:hamiltonian}
H(x,v,\tau; z) = \frac{v^2}{2} + V(x) + \mathcal{E}(\tau,z)x.
\end{equation}
We use the soft-Coulomb potential $V(x) = -(x^2+a^2)^{-1/2}$ to describe the electron-ion interaction.
For the quantum model, $\langle v(\tau,z) \rangle$ is obtained from the solution of the TDSE with Hamiltonian \eqref{eq:hamiltonian} at every $z$ \cite{Shon00}.
Details on the numerical schemes employed to solve our model equations are provided in the Supplemental Material \footnote{See Supplemental Material for an illustration of the model geometry, a description of the numerical schemes used to solve the model equations, and more details on the computation of the observables plotted in the figures, which includes Refs.~\cite{Evst13,Ruth83}.}.
%
\begin{figure}
\includegraphics[width=0.45\textwidth]{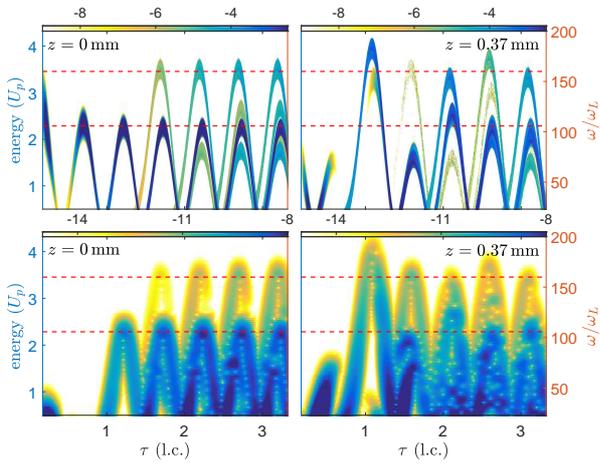}
\caption{Electron dynamics along different points of the propagation. Left panels: $z = 0\,\,\mathrm{mm}$, at the gas entrance. Right panels: $z = 0.37\,\,\mathrm{mm}$. Top panels: Kinetic energy distribution of recolliding electrons as a function of time, with the density indicated by a logarithmic color scale, from the classical model. Bottom panels: Time-frequency maps of the dipole acceleration $-\langle \partial V/ \partial x(\tau,z) \rangle$, computed with a time window of $T=0.35$ laser cycles, from the quantum model. The intensity of radiation at a given time and frequency is indicated by a logarithmic color scale. In all panels, the red dashed lines indicate the kinetic energies and corresponding frequencies of $2U_p+I_p$ and $3.17U_p+I_p$.}\label{fig:time-freq}
\end{figure}

On a single-atom level, the classical model does not capture the high-harmonic plateau and cutoff because it lacks quantum interference effects \cite{Sand99}.
While it may do a fair job for the low order harmonics originating from the nonlinear response of the bounded part of the wavepacket \cite{Band92,Uzdi10}, its spectrum is not accurate when the harmonics are driven by ionized, recolliding trajectories \cite{Sand99,Prot96,Both09}.
Evidently, this situation is unchanged when the classical single atoms are allowed to interact via the electromagnetic field, as shown in the spectrum of Fig.~\ref{fig:spectra}.
At the same time, the quantitative agreement between the classical and the quantum models for low frequencies persists during propagation.
Because these frequencies are the dominant constituents of the field (see Fig.~\ref{fig:spectra}), this suggests that the classical model provides a faithful representation of the true electron dynamics underlying the quantum model all along the propagation.

This expectation is indeed borne out by the excellent correspondence between statistics of electron encounters with the core, or recollisions, obtained from the classical model and the dipole radiation spectra obtained from the quantum model, displayed in Fig.~\ref{fig:time-freq}.
According to the semiclassical model \cite{Cork93,Kula93}, an electron which is driven to the ionic core with a kinetic energy $\kappa U_p$ may recombine into the atomic ground state with energy $-I_p$ and radiate its excess energy as a photon with frequency $\omega=\kappa U_p + I_p$.
This implies that a specific frequency $\omega$ is only emitted when an electron trajectory with the appropriate kinetic energy $\kappa U_p$ enters the core region \cite{Lewe94}, a behavior which can be revealed by performing a time-frequency analysis of the dipole acceleration $-\langle \partial V / \partial x (\tau,z) \rangle$ from the quantum model~\cite{Yako03,Pukh03}.
In the lower panels of Fig.~\ref{fig:time-freq}, the dipole acceleration time-frequency maps are displayed, showing which frequencies $\omega$ are generated at times $\tau$.
On the other hand, the top panels show the probability of an electron recollision with a given kinetic energy $\kappa U_p$ at time $\tau$, obtained from the classical model.
A strong correspondence between these two figures is expected on the left panels at $z=0$, when the classical and quantum atoms are driven by the exact same incident field, $\mathcal{E}(\tau,z=0)=E_0 \cos(\omega_L \tau)$.
However, after propagating to $z=0.37\,\mathrm{mm}$, the laser fields in each model have been driven by different dipole velocities.
Therefore, it is remarkable that the level of agreement between the two fields is so high that the dynamics of the electrons in both the classical and quantum cases continues to be very similar throughout propagation, evidenced by the continuing close correspondence between the two figures in the right panels.

The close correspondence of the two figures does not hold for all $\tau$, however.
For example, comparing the upper and lower left panels of Fig.~\ref{fig:time-freq}, we see that at $z=0$, there are recollisions which occur with $\kappa \approx 2$ during the first laser cycle in the classical model without emission of the corresponding high harmonics in the quantum model \cite{Prot96}. 
The reason is the total depopulation of the ground state \cite{Pukh03} at the beginning of the pulse, or more generally speaking the complete lack of an electron wavepacket at least transiently bounded to the ion.
Because the highest harmonic frequencies are generated by the quantum interference between a bounded wavepacket and a recolliding wavepacket \cite{Sand99}, a complete lack of a bounded wavepacket means very high harmonics cannot be emitted.
On the other hand, by comparing the upper and lower right panels of Fig.~\ref{fig:time-freq}, we see that by $z=0.37\,\mathrm{mm}$, there is high-harmonic recollision-driven radiation emitted during the first laser cycle.
This suggests the creation of a bound state earlier in the laser pulse as the propagation proceeds.
Here, we exploit the main advantage of the classical model: its ability to confirm and analyze this scenario by visualizing the electron dynamics in phase space.

\begin{figure}
\includegraphics[width=0.45\textwidth]{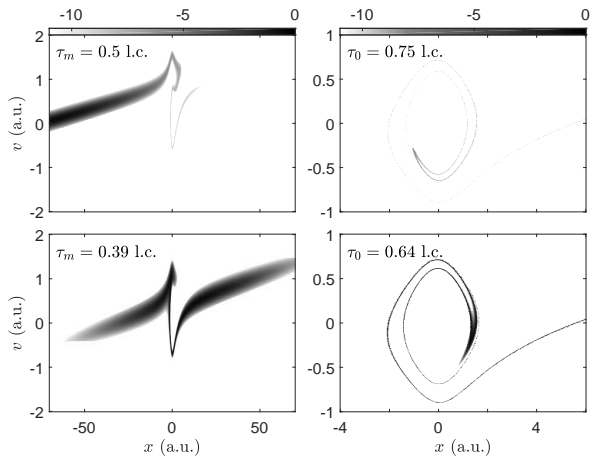}
\caption{Snapshots of the distribution function $f(x,v,\tau;z)$ in a logarithmic scale. Left panels: $f(x,v,\tau_m;z)$, where $\tau_m > 0$ is the first field intensity maximum after the start of the pulse. Right panels: $f(x,v,\tau_0;z)$, where $\tau_0 > \tau_m$ is the first zero of the field following $\tau_m$. The upper panels are at $z=0$, while the lower panels are at $z=0.37\,\,\mathrm{mm}$.}\label{fig:snaps}
\end{figure}

In Fig.~\ref{fig:snaps}, we show snapshots of the classical electron distribution function $f(x,v,\tau;z)$ at particular times $\tau$ and propagation positions $z$.
With the initial conditions we have chosen, the electron wavepacket always begins at rest far on the right side of the core, and the field is positive and at a maximum.
Therefore, the electron is always initially driven to the left, towards the core.
At $z=0$, the electron wavepacket is predominantly on the left of the ion by time $\tau_m$ when the laser field has reversed direction once and is again at an extremum.
When the field subsequently goes to zero at time $\tau_0$, the wavepacket has almost completely vacated the core region.
However, as propagation proceeds, the blueshift causes the laser field to reverse direction earlier in the pulse, and this causes the center of the wavepacket at time $\tau_m$ to be displaced to the right.
By $z=0.37\,\,\mathrm{mm}$, the wavepacket is thus nearly centered over the ion, with the electron velocities distributed about zero.
When part of the wavepacket arrives to the core with a low kinetic energy like this, it has a high probability of becoming trapped there \cite{Zago12}, and indeed a trapped part of the wavepacket is clearly observed in the subsequent snapshot of the distribution function at $\tau_0$ (lower right panel of Fig.~\ref{fig:snaps}).
This confirms that a bound state is created earlier in the pulse after propagation through part of the gas.
Furthermore, this explains the emergence of recollision-driven high-harmonic radiation for $\tau < 1\,\,\mathrm{l.c.}$ at longer propagation distances, despite this radiation being absent in this time interval at $z=0$.

In addition to providing an explanation of the electron trapping, the classical model also allows us to explain the anomalous high harmonic radiation in excess of the $3.17 U_p + I_p$ cutoff law that we observed in the quantum calculation.
At $z=0$, the left panels of Fig.~\ref{fig:time-freq} indicate that both the recollision kinetic energy and dipole radiation cutoffs are in the expected place, with the cutoff frequency being $\omega_c = 160\omega_L$ in this case.
Thus, for small $z$, there is no coherent growth of frequencies $\omega > \omega_c$, as shown in the lower inset of Fig.~\ref{fig:spectra}.
However, as propagation proceeds, a family of recollisions appears, the first blue arc in the upper right panel of Fig.~\ref{fig:time-freq}, that returns in the approximate range $0.75 < \tau < 1.25\,\,\mathrm{l.c.}$ and has a maximum kinetic energy of about $4.1 U_p$ or $190\omega_L$.
A trajectory analysis reveals that this new family of recollisions also has a low kinetic energy encounter with the core near $\tau_m$, just like the early trapped trajectories.
Rather than becoming trapped, however, these trajectories escape the core once more, returning with a range of kinetic energies up to about $4.1 U_p$. 
This energy agrees well with a calculation of the maximum return kinetic energy of a free electron in the field ${\cal E}(\tau,z)$ ionized at $\tau \sim \tau_m$, with the effect of the Coulomb potential treated as a perturbation \cite{Kamo14}.
Because this calculation predicts a cutoff of $3.17U_p + I_p$ for a monochromatic laser field, we conclude that the increase in kinetic energy above the usual high-harmonic cutoff is due to the departure of the field from a monochromatic wave throughout propagation.
It is these recollisions' radiation which drives the evolution in energy of the macroscopic electric field in the frequency band of $175\omega_L$ to $200\omega_L$ starting at $z=0.2\,\,\mathrm{mm}$, shown in the lower inset of Fig.~\ref{fig:spectra}.
The circumstances permitting the anomalous high harmonic radiation are maintained over a substantial propagation length, leading first to the coherent growth of the energy in these frequencies, followed by the coherent absorption beginning at about $z =  0.5\,\,\mathrm{mm}$.

In summary, we have presented a purely classical model for HHG during the propagation of intense laser pulses through atomic gases.
The model agrees quantitatively with a quantum model for the low frequency components of the laser field in the case of an initially monochromatic pulse propagating through a singly ionized gas, and the phase-space analysis permitted by the classical model explains the extension of the high-harmonic cutoff observed in the quantum simulation.
Our model may be useful in the analysis of experimentally-observed excessively high harmonic radiation, though it is rare in gases \cite{Vale04}, as well as the analysis and further refinement of trajectory-based semiclassical computational schemes \cite{Zago12,Higu11} and control strategies for HHG \cite{Briz13,Haes14}.
Other possible applications of our model include the study of terahertz emission from field-ionized gases, where a classical description of the electron motion is also germane \cite{Mart15}, and the study of filamentation, where first-principles descriptions of the atom-field interaction are increasingly sought \cite{Schu17}.

We acknowledge Fran\c{c}ois Mauger for helpful discussions. The project leading to this research has received funding from the European Union's Horizon 2020 research and innovation program under the Marie Sk{\l}odowska-Curie grant agreement No 734557. S.A.B. and T.U. acknowledge funding from the NSF (Grant No. PHY1602823). This material is based upon research supported by the Chateaubriand Fellowship of the Office for Science \& Technology of the Embassy of France in the United States.


\begin{thebibliography}{44}%
\makeatletter
\providecommand \@ifxundefined [1]{%
 \@ifx{#1\undefined}
}%
\providecommand \@ifnum [1]{%
 \ifnum #1\expandafter \@firstoftwo
 \else \expandafter \@secondoftwo
 \fi
}%
\providecommand \@ifx [1]{%
 \ifx #1\expandafter \@firstoftwo
 \else \expandafter \@secondoftwo
 \fi
}%
\providecommand \natexlab [1]{#1}%
\providecommand \enquote  [1]{``#1''}%
\providecommand \bibnamefont  [1]{#1}%
\providecommand \bibfnamefont [1]{#1}%
\providecommand \citenamefont [1]{#1}%
\providecommand \href@noop [0]{\@secondoftwo}%
\providecommand \href [0]{\begingroup \@sanitize@url \@href}%
\providecommand \@href[1]{\@@startlink{#1}\@@href}%
\providecommand \@@href[1]{\endgroup#1\@@endlink}%
\providecommand \@sanitize@url [0]{\catcode `\\12\catcode `\$12\catcode
  `\&12\catcode `\#12\catcode `\^12\catcode `\_12\catcode `\%12\relax}%
\providecommand \@@startlink[1]{}%
\providecommand \@@endlink[0]{}%
\providecommand \url  [0]{\begingroup\@sanitize@url \@url }%
\providecommand \@url [1]{\endgroup\@href {#1}{\urlprefix }}%
\providecommand \urlprefix  [0]{URL }%
\providecommand \Eprint [0]{\href }%
\providecommand \doibase [0]{http://dx.doi.org/}%
\providecommand \selectlanguage [0]{\@gobble}%
\providecommand \bibinfo  [0]{\@secondoftwo}%
\providecommand \bibfield  [0]{\@secondoftwo}%
\providecommand \translation [1]{[#1]}%
\providecommand \BibitemOpen [0]{}%
\providecommand \bibitemStop [0]{}%
\providecommand \bibitemNoStop [0]{.\EOS\space}%
\providecommand \EOS [0]{\spacefactor3000\relax}%
\providecommand \BibitemShut  [1]{\csname bibitem#1\endcsname}%
\let\auto@bib@innerbib\@empty
\bibitem [{\citenamefont {Haessler}\ \emph {et~al.}(2014)\citenamefont
  {Haessler}, \citenamefont {Bal{\v{c}}iunas}, \citenamefont {Fan},
  \citenamefont {Andriukaitis}, \citenamefont {Pug{\v{z}}lys}, \citenamefont
  {Baltu{\v{s}}ka}, \citenamefont {Witting}, \citenamefont {Squibb},
  \citenamefont {Za{\"\i}r}, \citenamefont {Tisch} \emph {et~al.}}]{Haes14}%
  \BibitemOpen
  \bibfield  {author} {\bibinfo {author} {\bibfnamefont {S.}~\bibnamefont
  {Haessler}}, \bibinfo {author} {\bibfnamefont {T.}~\bibnamefont
  {Bal{\v{c}}iunas}}, \bibinfo {author} {\bibfnamefont {G.}~\bibnamefont
  {Fan}}, \bibinfo {author} {\bibfnamefont {G.}~\bibnamefont {Andriukaitis}},
  \bibinfo {author} {\bibfnamefont {A.}~\bibnamefont {Pug{\v{z}}lys}}, \bibinfo
  {author} {\bibfnamefont {A.}~\bibnamefont {Baltu{\v{s}}ka}}, \bibinfo
  {author} {\bibfnamefont {T.}~\bibnamefont {Witting}}, \bibinfo {author}
  {\bibfnamefont {R.}~\bibnamefont {Squibb}}, \bibinfo {author} {\bibfnamefont
  {A.}~\bibnamefont {Za{\"\i}r}}, \bibinfo {author} {\bibfnamefont {J.~W.}\
  \bibnamefont {Tisch}},  \emph {et~al.},\ }\href@noop {} {\bibfield  {journal}
  {\bibinfo  {journal} {Phys.~Rev.~X}\ }\textbf {\bibinfo {volume} {4}},\
  \bibinfo {pages} {021028} (\bibinfo {year} {2014})}\BibitemShut {NoStop}%
\bibitem [{\citenamefont {Kazamias}\ \emph {et~al.}(2003)\citenamefont
  {Kazamias}, \citenamefont {Douillet}, \citenamefont {Weihe}, \citenamefont
  {Valentin}, \citenamefont {Rousse}, \citenamefont {Sebban}, \citenamefont
  {Grillon}, \citenamefont {Aug{\'e}}, \citenamefont {Hulin},\ and\
  \citenamefont {Balcou}}]{Kaza03}%
  \BibitemOpen
  \bibfield  {author} {\bibinfo {author} {\bibfnamefont {S.}~\bibnamefont
  {Kazamias}}, \bibinfo {author} {\bibfnamefont {D.}~\bibnamefont {Douillet}},
  \bibinfo {author} {\bibfnamefont {F.}~\bibnamefont {Weihe}}, \bibinfo
  {author} {\bibfnamefont {C.}~\bibnamefont {Valentin}}, \bibinfo {author}
  {\bibfnamefont {A.}~\bibnamefont {Rousse}}, \bibinfo {author} {\bibfnamefont
  {S.}~\bibnamefont {Sebban}}, \bibinfo {author} {\bibfnamefont
  {G.}~\bibnamefont {Grillon}}, \bibinfo {author} {\bibfnamefont
  {F.}~\bibnamefont {Aug{\'e}}}, \bibinfo {author} {\bibfnamefont
  {D.}~\bibnamefont {Hulin}}, \ and\ \bibinfo {author} {\bibfnamefont
  {P.}~\bibnamefont {Balcou}},\ }\href@noop {} {\bibfield  {journal} {\bibinfo
  {journal} {Phys.~Rev.~Lett.}\ }\textbf {\bibinfo {volume} {90}},\ \bibinfo
  {pages} {193901} (\bibinfo {year} {2003})}\BibitemShut {NoStop}%
\bibitem [{\citenamefont {Constant}\ \emph {et~al.}(1999)\citenamefont
  {Constant}, \citenamefont {Garzella}, \citenamefont {Breger}, \citenamefont
  {M\'evel}, \citenamefont {Dorrer}, \citenamefont {Le~Blanc}, \citenamefont
  {Salin},\ and\ \citenamefont {Agostini}}]{Cons99}%
  \BibitemOpen
  \bibfield  {author} {\bibinfo {author} {\bibfnamefont {E.}~\bibnamefont
  {Constant}}, \bibinfo {author} {\bibfnamefont {D.}~\bibnamefont {Garzella}},
  \bibinfo {author} {\bibfnamefont {P.}~\bibnamefont {Breger}}, \bibinfo
  {author} {\bibfnamefont {E.}~\bibnamefont {M\'evel}}, \bibinfo {author}
  {\bibfnamefont {C.}~\bibnamefont {Dorrer}}, \bibinfo {author} {\bibfnamefont
  {C.}~\bibnamefont {Le~Blanc}}, \bibinfo {author} {\bibfnamefont
  {F.}~\bibnamefont {Salin}}, \ and\ \bibinfo {author} {\bibfnamefont
  {P.}~\bibnamefont {Agostini}},\ }\href {\doibase 10.1103/PhysRevLett.82.1668}
  {\bibfield  {journal} {\bibinfo  {journal} {Phys. Rev. Lett.}\ }\textbf
  {\bibinfo {volume} {82}},\ \bibinfo {pages} {1668} (\bibinfo {year}
  {1999})}\BibitemShut {NoStop}%
\bibitem [{\citenamefont {Popmintchev}\ \emph {et~al.}(2012)\citenamefont
  {Popmintchev}, \citenamefont {Chen}, \citenamefont {Popmintchev},
  \citenamefont {Arpin}, \citenamefont {Brown}, \citenamefont {Ali{\v
  s}auskas}, \citenamefont {Andriukaitis}, \citenamefont {Bal{\v c}iunas},
  \citenamefont {M{\"u}cke}, \citenamefont {Pugzlys}, \citenamefont {Baltu{\v
  s}ka}, \citenamefont {Shim}, \citenamefont {Schrauth}, \citenamefont {Gaeta},
  \citenamefont {Hern{\'a}ndez-Garc{\'\i}a}, \citenamefont {Plaja},
  \citenamefont {Becker}, \citenamefont {Jaron-Becker}, \citenamefont
  {Murnane},\ and\ \citenamefont {Kapteyn}}]{Popm12}%
  \BibitemOpen
  \bibfield  {author} {\bibinfo {author} {\bibfnamefont {T.}~\bibnamefont
  {Popmintchev}}, \bibinfo {author} {\bibfnamefont {M.-C.}\ \bibnamefont
  {Chen}}, \bibinfo {author} {\bibfnamefont {D.}~\bibnamefont {Popmintchev}},
  \bibinfo {author} {\bibfnamefont {P.}~\bibnamefont {Arpin}}, \bibinfo
  {author} {\bibfnamefont {S.}~\bibnamefont {Brown}}, \bibinfo {author}
  {\bibfnamefont {S.}~\bibnamefont {Ali{\v s}auskas}}, \bibinfo {author}
  {\bibfnamefont {G.}~\bibnamefont {Andriukaitis}}, \bibinfo {author}
  {\bibfnamefont {T.}~\bibnamefont {Bal{\v c}iunas}}, \bibinfo {author}
  {\bibfnamefont {O.}~\bibnamefont {M{\"u}cke}}, \bibinfo {author}
  {\bibfnamefont {A.}~\bibnamefont {Pugzlys}}, \bibinfo {author} {\bibfnamefont
  {A.}~\bibnamefont {Baltu{\v s}ka}}, \bibinfo {author} {\bibfnamefont
  {B.}~\bibnamefont {Shim}}, \bibinfo {author} {\bibfnamefont {S.}~\bibnamefont
  {Schrauth}}, \bibinfo {author} {\bibfnamefont {A.}~\bibnamefont {Gaeta}},
  \bibinfo {author} {\bibfnamefont {C.}~\bibnamefont
  {Hern{\'a}ndez-Garc{\'\i}a}}, \bibinfo {author} {\bibfnamefont
  {L.}~\bibnamefont {Plaja}}, \bibinfo {author} {\bibfnamefont
  {A.}~\bibnamefont {Becker}}, \bibinfo {author} {\bibfnamefont
  {A.}~\bibnamefont {Jaron-Becker}}, \bibinfo {author} {\bibfnamefont
  {M.}~\bibnamefont {Murnane}}, \ and\ \bibinfo {author} {\bibfnamefont
  {H.}~\bibnamefont {Kapteyn}},\ }\href {\doibase 10.1126/science.1218497}
  {\bibfield  {journal} {\bibinfo  {journal} {Science}\ }\textbf {\bibinfo
  {volume} {336}},\ \bibinfo {pages} {1287} (\bibinfo {year}
  {2012})}\BibitemShut {NoStop}%
\bibitem [{\citenamefont {Popmintchev}\ \emph {et~al.}(2015)\citenamefont
  {Popmintchev}, \citenamefont {Hern{\'a}ndez-Garc{\'\i}a}, \citenamefont
  {Dollar}, \citenamefont {Mancuso}, \citenamefont {P{\'e}rez-Hern{\'a}ndez},
  \citenamefont {Chen}, \citenamefont {Hankla}, \citenamefont {Gao},
  \citenamefont {Shim}, \citenamefont {Gaeta} \emph {et~al.}}]{Popm15}%
  \BibitemOpen
  \bibfield  {author} {\bibinfo {author} {\bibfnamefont {D.}~\bibnamefont
  {Popmintchev}}, \bibinfo {author} {\bibfnamefont {C.}~\bibnamefont
  {Hern{\'a}ndez-Garc{\'\i}a}}, \bibinfo {author} {\bibfnamefont
  {F.}~\bibnamefont {Dollar}}, \bibinfo {author} {\bibfnamefont
  {C.}~\bibnamefont {Mancuso}}, \bibinfo {author} {\bibfnamefont {J.~A.}\
  \bibnamefont {P{\'e}rez-Hern{\'a}ndez}}, \bibinfo {author} {\bibfnamefont
  {M.-C.}\ \bibnamefont {Chen}}, \bibinfo {author} {\bibfnamefont
  {A.}~\bibnamefont {Hankla}}, \bibinfo {author} {\bibfnamefont
  {X.}~\bibnamefont {Gao}}, \bibinfo {author} {\bibfnamefont {B.}~\bibnamefont
  {Shim}}, \bibinfo {author} {\bibfnamefont {A.~L.}\ \bibnamefont {Gaeta}},
  \emph {et~al.},\ }\href@noop {} {\bibfield  {journal} {\bibinfo  {journal}
  {Science}\ }\textbf {\bibinfo {volume} {350}},\ \bibinfo {pages} {1225}
  (\bibinfo {year} {2015})}\BibitemShut {NoStop}%
\bibitem [{\citenamefont {Rae}\ \emph {et~al.}(1994)\citenamefont {Rae},
  \citenamefont {Burnett},\ and\ \citenamefont {Cooper}}]{Rae94}%
  \BibitemOpen
  \bibfield  {author} {\bibinfo {author} {\bibfnamefont {S.~C.}\ \bibnamefont
  {Rae}}, \bibinfo {author} {\bibfnamefont {K.}~\bibnamefont {Burnett}}, \ and\
  \bibinfo {author} {\bibfnamefont {J.}~\bibnamefont {Cooper}},\ }\href@noop {}
  {\bibfield  {journal} {\bibinfo  {journal} {Phys.~Rev.~A}\ }\textbf {\bibinfo
  {volume} {50}},\ \bibinfo {pages} {3438} (\bibinfo {year}
  {1994})}\BibitemShut {NoStop}%
\bibitem [{\citenamefont {Geissler}\ \emph {et~al.}(1999)\citenamefont
  {Geissler}, \citenamefont {Tempea}, \citenamefont {Scrinzi}, \citenamefont
  {Schn{\"u}rer}, \citenamefont {Krausz},\ and\ \citenamefont
  {Brabec}}]{Geis99}%
  \BibitemOpen
  \bibfield  {author} {\bibinfo {author} {\bibfnamefont {M.}~\bibnamefont
  {Geissler}}, \bibinfo {author} {\bibfnamefont {G.}~\bibnamefont {Tempea}},
  \bibinfo {author} {\bibfnamefont {A.}~\bibnamefont {Scrinzi}}, \bibinfo
  {author} {\bibfnamefont {M.}~\bibnamefont {Schn{\"u}rer}}, \bibinfo {author}
  {\bibfnamefont {F.}~\bibnamefont {Krausz}}, \ and\ \bibinfo {author}
  {\bibfnamefont {T.}~\bibnamefont {Brabec}},\ }\href@noop {} {\bibfield
  {journal} {\bibinfo  {journal} {Phys.~Rev.~Lett.}\ }\textbf {\bibinfo
  {volume} {83}},\ \bibinfo {pages} {2930} (\bibinfo {year}
  {1999})}\BibitemShut {NoStop}%
\bibitem [{\citenamefont {Gaarde}\ \emph {et~al.}(2006)\citenamefont {Gaarde},
  \citenamefont {Murakami},\ and\ \citenamefont {Kienberger}}]{Gaar06}%
  \BibitemOpen
  \bibfield  {author} {\bibinfo {author} {\bibfnamefont {M.~B.}\ \bibnamefont
  {Gaarde}}, \bibinfo {author} {\bibfnamefont {M.}~\bibnamefont {Murakami}}, \
  and\ \bibinfo {author} {\bibfnamefont {R.}~\bibnamefont {Kienberger}},\
  }\href {\doibase 10.1103/PhysRevA.74.053401} {\bibfield  {journal} {\bibinfo
  {journal} {Phys. Rev. A}\ }\textbf {\bibinfo {volume} {74}},\ \bibinfo
  {pages} {053401} (\bibinfo {year} {2006})}\BibitemShut {NoStop}%
\bibitem [{\citenamefont {Gaarde}\ \emph {et~al.}(2008)\citenamefont {Gaarde},
  \citenamefont {Tate},\ and\ \citenamefont {Schafer}}]{Gaar08}%
  \BibitemOpen
  \bibfield  {author} {\bibinfo {author} {\bibfnamefont {M.~B.}\ \bibnamefont
  {Gaarde}}, \bibinfo {author} {\bibfnamefont {J.~L.}\ \bibnamefont {Tate}}, \
  and\ \bibinfo {author} {\bibfnamefont {K.~J.}\ \bibnamefont {Schafer}},\
  }\href {http://stacks.iop.org/0953-4075/41/i=13/a=132001} {\bibfield
  {journal} {\bibinfo  {journal} {J.~Phys.~B}\ }\textbf {\bibinfo {volume}
  {41}},\ \bibinfo {pages} {132001} (\bibinfo {year} {2008})}\BibitemShut
  {NoStop}%
\bibitem [{\citenamefont {Popmintchev}\ \emph {et~al.}(2010)\citenamefont
  {Popmintchev}, \citenamefont {Chen}, \citenamefont {Arpin}, \citenamefont
  {Murnane},\ and\ \citenamefont {Kapteyn}}]{Popm10}%
  \BibitemOpen
  \bibfield  {author} {\bibinfo {author} {\bibfnamefont {T.}~\bibnamefont
  {Popmintchev}}, \bibinfo {author} {\bibfnamefont {M.-C.}\ \bibnamefont
  {Chen}}, \bibinfo {author} {\bibfnamefont {P.}~\bibnamefont {Arpin}},
  \bibinfo {author} {\bibfnamefont {M.~M.}\ \bibnamefont {Murnane}}, \ and\
  \bibinfo {author} {\bibfnamefont {H.~C.}\ \bibnamefont {Kapteyn}},\
  }\href@noop {} {\bibfield  {journal} {\bibinfo  {journal} {Nat. Photonics}\
  }\textbf {\bibinfo {volume} {4}},\ \bibinfo {pages} {822} (\bibinfo {year}
  {2010})}\BibitemShut {NoStop}%
\bibitem [{\citenamefont {Lorin}\ and\ \citenamefont
  {Bandrauk}(2012)}]{Lori12}%
  \BibitemOpen
  \bibfield  {author} {\bibinfo {author} {\bibfnamefont {E.}~\bibnamefont
  {Lorin}}\ and\ \bibinfo {author} {\bibfnamefont {A.~D.}\ \bibnamefont
  {Bandrauk}},\ }\href@noop {} {\bibfield  {journal} {\bibinfo  {journal} {J.
  Comput. Sci.}\ }\textbf {\bibinfo {volume} {3}},\ \bibinfo {pages} {159}
  (\bibinfo {year} {2012})}\BibitemShut {NoStop}%
\bibitem [{\citenamefont {Gaarde}\ \emph {et~al.}(2011)\citenamefont {Gaarde},
  \citenamefont {Buth}, \citenamefont {Tate},\ and\ \citenamefont
  {Schafer}}]{Gaar11}%
  \BibitemOpen
  \bibfield  {author} {\bibinfo {author} {\bibfnamefont {M.~B.}\ \bibnamefont
  {Gaarde}}, \bibinfo {author} {\bibfnamefont {C.}~\bibnamefont {Buth}},
  \bibinfo {author} {\bibfnamefont {J.~L.}\ \bibnamefont {Tate}}, \ and\
  \bibinfo {author} {\bibfnamefont {K.~J.}\ \bibnamefont {Schafer}},\
  }\href@noop {} {\bibfield  {journal} {\bibinfo  {journal} {Phys.~Rev.~A}\
  }\textbf {\bibinfo {volume} {83}},\ \bibinfo {pages} {013419} (\bibinfo
  {year} {2011})}\BibitemShut {NoStop}%
\bibitem [{\citenamefont {Brabec}\ and\ \citenamefont {Krausz}(2000)}]{Brab00}%
  \BibitemOpen
  \bibfield  {author} {\bibinfo {author} {\bibfnamefont {T.}~\bibnamefont
  {Brabec}}\ and\ \bibinfo {author} {\bibfnamefont {F.}~\bibnamefont
  {Krausz}},\ }\href@noop {} {\bibfield  {journal} {\bibinfo  {journal}
  {Rev.~Mod.~Phys.}\ }\textbf {\bibinfo {volume} {72}},\ \bibinfo {pages} {545}
  (\bibinfo {year} {2000})}\BibitemShut {NoStop}%
\bibitem [{\citenamefont {Milosevic}\ \emph {et~al.}(2002)\citenamefont
  {Milosevic}, \citenamefont {Scrinzi},\ and\ \citenamefont {Brabec}}]{Milo02}%
  \BibitemOpen
  \bibfield  {author} {\bibinfo {author} {\bibfnamefont {N.}~\bibnamefont
  {Milosevic}}, \bibinfo {author} {\bibfnamefont {A.}~\bibnamefont {Scrinzi}},
  \ and\ \bibinfo {author} {\bibfnamefont {T.}~\bibnamefont {Brabec}},\
  }\href@noop {} {\bibfield  {journal} {\bibinfo  {journal} {Phys.~Rev.~Lett.}\
  }\textbf {\bibinfo {volume} {88}},\ \bibinfo {pages} {093905} (\bibinfo
  {year} {2002})}\BibitemShut {NoStop}%
\bibitem [{\citenamefont {Kulander}\ \emph {et~al.}(1993)\citenamefont
  {Kulander}, \citenamefont {Schafer},\ and\ \citenamefont {Krause}}]{Kula93}%
  \BibitemOpen
  \bibfield  {author} {\bibinfo {author} {\bibfnamefont {K.}~\bibnamefont
  {Kulander}}, \bibinfo {author} {\bibfnamefont {K.}~\bibnamefont {Schafer}}, \
  and\ \bibinfo {author} {\bibfnamefont {J.}~\bibnamefont {Krause}},\ }in\
  \href@noop {} {\emph {\bibinfo {booktitle} {Super-intense laser-atom
  physics}}}\ (\bibinfo  {publisher} {Springer},\ \bibinfo {year} {1993})\ pp.\
  \bibinfo {pages} {95--110}\BibitemShut {NoStop}%
\bibitem [{\citenamefont {Corkum}(1993)}]{Cork93}%
  \BibitemOpen
  \bibfield  {author} {\bibinfo {author} {\bibfnamefont {P.~B.}\ \bibnamefont
  {Corkum}},\ }\href@noop {} {\bibfield  {journal} {\bibinfo  {journal}
  {Phys.~Rev.~Lett.}\ }\textbf {\bibinfo {volume} {71}},\ \bibinfo {pages}
  {1994} (\bibinfo {year} {1993})}\BibitemShut {NoStop}%
\bibitem [{\citenamefont {Lewenstein}\ \emph {et~al.}(1994)\citenamefont
  {Lewenstein}, \citenamefont {Balcou}, \citenamefont {Ivanov}, \citenamefont
  {L'Huillier},\ and\ \citenamefont {Corkum}}]{Lewe94}%
  \BibitemOpen
  \bibfield  {author} {\bibinfo {author} {\bibfnamefont {M.}~\bibnamefont
  {Lewenstein}}, \bibinfo {author} {\bibfnamefont {P.}~\bibnamefont {Balcou}},
  \bibinfo {author} {\bibfnamefont {M.~Y.}\ \bibnamefont {Ivanov}}, \bibinfo
  {author} {\bibfnamefont {A.}~\bibnamefont {L'Huillier}}, \ and\ \bibinfo
  {author} {\bibfnamefont {P.~B.}\ \bibnamefont {Corkum}},\ }\href@noop {}
  {\bibfield  {journal} {\bibinfo  {journal} {Phys.~Rev.~A}\ }\textbf {\bibinfo
  {volume} {49}},\ \bibinfo {pages} {2117} (\bibinfo {year}
  {1994})}\BibitemShut {NoStop}%
\bibitem [{\citenamefont {Sali{\`e}res}\ \emph {et~al.}(2001)\citenamefont
  {Sali{\`e}res}, \citenamefont {Carr{\'e}}, \citenamefont {Le~D{\'e}roff},
  \citenamefont {Grasbon}, \citenamefont {Paulus}, \citenamefont {Walther},
  \citenamefont {Kopold}, \citenamefont {Becker}, \citenamefont
  {Milo{\v{s}}evi{\'c}}, \citenamefont {Sanpera} \emph {et~al.}}]{Sali01}%
  \BibitemOpen
  \bibfield  {author} {\bibinfo {author} {\bibfnamefont {P.}~\bibnamefont
  {Sali{\`e}res}}, \bibinfo {author} {\bibfnamefont {B.}~\bibnamefont
  {Carr{\'e}}}, \bibinfo {author} {\bibfnamefont {L.}~\bibnamefont
  {Le~D{\'e}roff}}, \bibinfo {author} {\bibfnamefont {F.}~\bibnamefont
  {Grasbon}}, \bibinfo {author} {\bibfnamefont {G.}~\bibnamefont {Paulus}},
  \bibinfo {author} {\bibfnamefont {H.}~\bibnamefont {Walther}}, \bibinfo
  {author} {\bibfnamefont {R.}~\bibnamefont {Kopold}}, \bibinfo {author}
  {\bibfnamefont {W.}~\bibnamefont {Becker}}, \bibinfo {author} {\bibfnamefont
  {D.}~\bibnamefont {Milo{\v{s}}evi{\'c}}}, \bibinfo {author} {\bibfnamefont
  {A.}~\bibnamefont {Sanpera}},  \emph {et~al.},\ }\href@noop {} {\bibfield
  {journal} {\bibinfo  {journal} {Science}\ }\textbf {\bibinfo {volume}
  {292}},\ \bibinfo {pages} {902} (\bibinfo {year} {2001})}\BibitemShut
  {NoStop}%
\bibitem [{\citenamefont {Chipperfield}\ \emph {et~al.}(2009)\citenamefont
  {Chipperfield}, \citenamefont {Robinson}, \citenamefont {Tisch},\ and\
  \citenamefont {Marangos}}]{Chip09}%
  \BibitemOpen
  \bibfield  {author} {\bibinfo {author} {\bibfnamefont {L.~E.}\ \bibnamefont
  {Chipperfield}}, \bibinfo {author} {\bibfnamefont {J.~S.}\ \bibnamefont
  {Robinson}}, \bibinfo {author} {\bibfnamefont {J.~W.~G.}\ \bibnamefont {Tisch}},
  \ and\ \bibinfo {author} {\bibfnamefont {J.~P.}\ \bibnamefont {Marangos}},\
  }\href@noop {} {\bibfield  {journal} {\bibinfo  {journal} {Phys.~Rev.~Lett.}\
  }\textbf {\bibinfo {volume} {102}},\ \bibinfo {pages} {063003} (\bibinfo
  {year} {2009})}\BibitemShut {NoStop}%
\bibitem [{\citenamefont {Brizuela}\ \emph {et~al.}(2013)\citenamefont
  {Brizuela}, \citenamefont {Heyl}, \citenamefont {Rudawski}, \citenamefont
  {Kroon}, \citenamefont {Rading}, \citenamefont {Dahlstr{\"o}m}, \citenamefont
  {Mauritsson}, \citenamefont {Johnsson}, \citenamefont {Arnold},\ and\
  \citenamefont {L'Huillier}}]{Briz13}%
  \BibitemOpen
  \bibfield  {author} {\bibinfo {author} {\bibfnamefont {F.}~\bibnamefont
  {Brizuela}}, \bibinfo {author} {\bibfnamefont {C.~M.}\ \bibnamefont {Heyl}},
  \bibinfo {author} {\bibfnamefont {P.}~\bibnamefont {Rudawski}}, \bibinfo
  {author} {\bibfnamefont {D.}~\bibnamefont {Kroon}}, \bibinfo {author}
  {\bibfnamefont {L.}~\bibnamefont {Rading}}, \bibinfo {author} {\bibfnamefont
  {J.~M.}\ \bibnamefont {Dahlstr{\"o}m}}, \bibinfo {author} {\bibfnamefont
  {J.}~\bibnamefont {Mauritsson}}, \bibinfo {author} {\bibfnamefont
  {P.}~\bibnamefont {Johnsson}}, \bibinfo {author} {\bibfnamefont {C.~L.}\
  \bibnamefont {Arnold}}, \ and\ \bibinfo {author} {\bibfnamefont
  {A.}~\bibnamefont {L'Huillier}},\ }\href@noop {} {\bibfield  {journal}
  {\bibinfo  {journal} {Sci. Rep.}\ }\textbf {\bibinfo {volume} {3}} (\bibinfo
  {year} {2013})}\BibitemShut {NoStop}%
\bibitem [{\citenamefont {Yost}\ \emph {et~al.}(2009)\citenamefont {Yost},
  \citenamefont {Schibli}, \citenamefont {Ye}, \citenamefont {Tate},
  \citenamefont {Hostetter}, \citenamefont {Gaarde},\ and\ \citenamefont
  {Schafer}}]{Yost09}%
  \BibitemOpen
  \bibfield  {author} {\bibinfo {author} {\bibfnamefont {D.~C.}\ \bibnamefont
  {Yost}}, \bibinfo {author} {\bibfnamefont {T.~R.}\ \bibnamefont {Schibli}},
  \bibinfo {author} {\bibfnamefont {J.}~\bibnamefont {Ye}}, \bibinfo {author}
  {\bibfnamefont {J.~L.}\ \bibnamefont {Tate}}, \bibinfo {author}
  {\bibfnamefont {J.}~\bibnamefont {Hostetter}}, \bibinfo {author}
  {\bibfnamefont {M.~B.}\ \bibnamefont {Gaarde}}, \ and\ \bibinfo {author}
  {\bibfnamefont {K.~J.}\ \bibnamefont {Schafer}},\ }\href {\doibase
  10.1038/nphys1398} {\bibfield  {journal} {\bibinfo  {journal} {Nat. Phys.}\
  }\textbf {\bibinfo {volume} {5}},\ \bibinfo {pages} {815} (\bibinfo {year}
  {2009})}\BibitemShut {NoStop}%
\bibitem [{\citenamefont {Xiong}\ \emph {et~al.}(2014)\citenamefont {Xiong},
  \citenamefont {Geng}, \citenamefont {Tang}, \citenamefont {Peng},\ and\
  \citenamefont {Gong}}]{Xion14}%
  \BibitemOpen
  \bibfield  {author} {\bibinfo {author} {\bibfnamefont {W.-H.}\ \bibnamefont
  {Xiong}}, \bibinfo {author} {\bibfnamefont {J.-W.}\ \bibnamefont {Geng}},
  \bibinfo {author} {\bibfnamefont {J.-Y.}\ \bibnamefont {Tang}}, \bibinfo
  {author} {\bibfnamefont {L.-Y.}\ \bibnamefont {Peng}}, \ and\ \bibinfo
  {author} {\bibfnamefont {Q.}~\bibnamefont {Gong}},\ }\href {\doibase
  10.1103/PhysRevLett.112.233001} {\bibfield  {journal} {\bibinfo  {journal}
  {Phys. Rev. Lett.}\ }\textbf {\bibinfo {volume} {112}},\ \bibinfo {pages}
  {233001} (\bibinfo {year} {2014})}\BibitemShut {NoStop}%
\bibitem [{\citenamefont {Xiong}\ \emph {et~al.}(2017)\citenamefont {Xiong},
  \citenamefont {Peng},\ and\ \citenamefont {Gong}}]{Xion17}%
  \BibitemOpen
  \bibfield  {author} {\bibinfo {author} {\bibfnamefont {W.-H.}\ \bibnamefont
  {Xiong}}, \bibinfo {author} {\bibfnamefont {L.-Y.}\ \bibnamefont {Peng}}, \
  and\ \bibinfo {author} {\bibfnamefont {Q.}~\bibnamefont {Gong}},\ }\href@noop
  {} {\bibfield  {journal} {\bibinfo  {journal} {J.~Phys.~B}\ }\textbf
  {\bibinfo {volume} {50}},\ \bibinfo {pages} {032001} (\bibinfo {year}
  {2017})}\BibitemShut {NoStop}%
\bibitem [{\citenamefont {Gaarde}\ \emph {et~al.}(2005)\citenamefont {Gaarde},
  \citenamefont {Schafer}, \citenamefont {Heinrich}, \citenamefont {Biegert},\
  and\ \citenamefont {Keller}}]{Gaar05}%
  \BibitemOpen
  \bibfield  {author} {\bibinfo {author} {\bibfnamefont {M.~B.}\ \bibnamefont
  {Gaarde}}, \bibinfo {author} {\bibfnamefont {K.~J.}\ \bibnamefont {Schafer}},
  \bibinfo {author} {\bibfnamefont {A.}~\bibnamefont {Heinrich}}, \bibinfo
  {author} {\bibfnamefont {J.}~\bibnamefont {Biegert}}, \ and\ \bibinfo
  {author} {\bibfnamefont {U.}~\bibnamefont {Keller}},\ }\href@noop {}
  {\bibfield  {journal} {\bibinfo  {journal} {Phys.~Rev.~A}\ }\textbf {\bibinfo
  {volume} {72}},\ \bibinfo {pages} {013411} (\bibinfo {year}
  {2005})}\BibitemShut {NoStop}%
\bibitem [{\citenamefont {Shafir}\ \emph {et~al.}(2012)\citenamefont {Shafir},
  \citenamefont {Fabre}, \citenamefont {Higuet}, \citenamefont {Soifer},
  \citenamefont {Dagan}, \citenamefont {Descamps}, \citenamefont {M\'evel},
  \citenamefont {Petit}, \citenamefont {W\"orner}, \citenamefont {Pons},
  \citenamefont {Dudovich},\ and\ \citenamefont {Mairesse}}]{Shaf12}%
  \BibitemOpen
  \bibfield  {author} {\bibinfo {author} {\bibfnamefont {D.}~\bibnamefont
  {Shafir}}, \bibinfo {author} {\bibfnamefont {B.}~\bibnamefont {Fabre}},
  \bibinfo {author} {\bibfnamefont {J.}~\bibnamefont {Higuet}}, \bibinfo
  {author} {\bibfnamefont {H.}~\bibnamefont {Soifer}}, \bibinfo {author}
  {\bibfnamefont {M.}~\bibnamefont {Dagan}}, \bibinfo {author} {\bibfnamefont
  {D.}~\bibnamefont {Descamps}}, \bibinfo {author} {\bibfnamefont
  {E.}~\bibnamefont {M\'evel}}, \bibinfo {author} {\bibfnamefont
  {S.}~\bibnamefont {Petit}}, \bibinfo {author} {\bibfnamefont {H.~J.}\
  \bibnamefont {W\"orner}}, \bibinfo {author} {\bibfnamefont {B.}~\bibnamefont
  {Pons}}, \bibinfo {author} {\bibfnamefont {N.}~\bibnamefont {Dudovich}}, \
  and\ \bibinfo {author} {\bibfnamefont {Y.}~\bibnamefont {Mairesse}},\
  }\href@noop {} {\bibfield  {journal} {\bibinfo  {journal} {Phys.~Rev.~Lett.}\
  }\textbf {\bibinfo {volume} {108}},\ \bibinfo {pages} {203001} (\bibinfo
  {year} {2012})}\BibitemShut {NoStop}%
\bibitem [{\citenamefont {Abanador}\ \emph {et~al.}(2017)\citenamefont
  {Abanador}, \citenamefont {Mauger}, \citenamefont {Lopata}, \citenamefont
  {Gaarde},\ and\ \citenamefont {Schafer}}]{Aban17}%
  \BibitemOpen
  \bibfield  {author} {\bibinfo {author} {\bibfnamefont {P.}~\bibnamefont
  {Abanador}}, \bibinfo {author} {\bibfnamefont {F.}~\bibnamefont {Mauger}},
  \bibinfo {author} {\bibfnamefont {K.}~\bibnamefont {Lopata}}, \bibinfo
  {author} {\bibfnamefont {M.}~\bibnamefont {Gaarde}}, \ and\ \bibinfo {author}
  {\bibfnamefont {K.}~\bibnamefont {Schafer}},\ }\href@noop {} {\bibfield
  {journal} {\bibinfo  {journal} {J.~Phys.~B}\ }\textbf {\bibinfo {volume}
  {50}},\ \bibinfo {pages} {035601} (\bibinfo {year} {2017})}\BibitemShut
  {NoStop}%
\bibitem [{\citenamefont {Protopapas}\ \emph {et~al.}(1996)\citenamefont
  {Protopapas}, \citenamefont {Lappas}, \citenamefont {Keitel},\ and\
  \citenamefont {Knight}}]{Prot96}%
  \BibitemOpen
  \bibfield  {author} {\bibinfo {author} {\bibfnamefont {M.}~\bibnamefont
  {Protopapas}}, \bibinfo {author} {\bibfnamefont {D.~G.}\ \bibnamefont
  {Lappas}}, \bibinfo {author} {\bibfnamefont {C.~H.}\ \bibnamefont {Keitel}},
  \ and\ \bibinfo {author} {\bibfnamefont {P.~L.}\ \bibnamefont {Knight}},\
  }\href {\doibase 10.1103/PhysRevA.53.R2933} {\bibfield  {journal} {\bibinfo
  {journal} {Phys. Rev. A}\ }\textbf {\bibinfo {volume} {53}},\ \bibinfo
  {pages} {R2933} (\bibinfo {year} {1996})}\BibitemShut {NoStop}%
\bibitem [{\citenamefont {van~de Sand}\ and\ \citenamefont
  {Rost}(1999)}]{Sand99}%
  \BibitemOpen
  \bibfield  {author} {\bibinfo {author} {\bibfnamefont {G.}~\bibnamefont
  {van~de Sand}}\ and\ \bibinfo {author} {\bibfnamefont {J.~M.}\ \bibnamefont
  {Rost}},\ }\href@noop {} {\bibfield  {journal} {\bibinfo  {journal}
  {Phys.~Rev.~Lett.}\ }\textbf {\bibinfo {volume} {83}},\ \bibinfo {pages}
  {524} (\bibinfo {year} {1999})}\BibitemShut {NoStop}%
\bibitem [{\citenamefont {Kim}\ and\ \citenamefont {Nam}(2002)}]{Kim02}%
  \BibitemOpen
  \bibfield  {author} {\bibinfo {author} {\bibfnamefont {J.-H.}\ \bibnamefont
  {Kim}}\ and\ \bibinfo {author} {\bibfnamefont {C.~H.}\ \bibnamefont {Nam}},\
  }\href@noop {} {\bibfield  {journal} {\bibinfo  {journal} {Phys.~Rev.~A}\
  }\textbf {\bibinfo {volume} {65}},\ \bibinfo {pages} {033801} (\bibinfo
  {year} {2002})}\BibitemShut {NoStop}%
\bibitem [{\citenamefont {Shon}\ \emph {et~al.}(2000)\citenamefont {Shon},
  \citenamefont {Suda},\ and\ \citenamefont {Midorikawa}}]{Shon00}%
  \BibitemOpen
  \bibfield  {author} {\bibinfo {author} {\bibfnamefont {N.~H.}\ \bibnamefont
  {Shon}}, \bibinfo {author} {\bibfnamefont {A.}~\bibnamefont {Suda}}, \ and\
  \bibinfo {author} {\bibfnamefont {K.}~\bibnamefont {Midorikawa}},\
  }\href@noop {} {\bibfield  {journal} {\bibinfo  {journal} {Phys.~Rev.~A}\
  }\textbf {\bibinfo {volume} {62}},\ \bibinfo {pages} {023801} (\bibinfo
  {year} {2000})}\BibitemShut {NoStop}%
\bibitem [{Note1()}]{Note1}%
  \BibitemOpen
  \bibinfo {note} {See Supplemental Material for an illustration of the model
  geometry, a description of the numerical schemes used to solve the model
  equations, and more details on the computation of the observables plotted in
  the figures, which includes Refs.~\cite {Evst13,Ruth83}.}\BibitemShut {Stop}%
\bibitem [{\citenamefont {Bandarage}\ \emph {et~al.}(1992)\citenamefont
  {Bandarage}, \citenamefont {Maquet}, \citenamefont {M\'enis}, \citenamefont
  {Ta\"{\i}eb}, \citenamefont {V\'eniard},\ and\ \citenamefont
  {Cooper}}]{Band92}%
  \BibitemOpen
  \bibfield  {author} {\bibinfo {author} {\bibfnamefont {G.}~\bibnamefont
  {Bandarage}}, \bibinfo {author} {\bibfnamefont {A.}~\bibnamefont {Maquet}},
  \bibinfo {author} {\bibfnamefont {T.}~\bibnamefont {M\'enis}}, \bibinfo
  {author} {\bibfnamefont {R.}~\bibnamefont {Ta\"{\i}eb}}, \bibinfo {author}
  {\bibfnamefont {V.}~\bibnamefont {V\'eniard}}, \ and\ \bibinfo {author}
  {\bibfnamefont {J.}~\bibnamefont {Cooper}},\ }\href {\doibase
  10.1103/PhysRevA.46.380} {\bibfield  {journal} {\bibinfo  {journal} {Phys.
  Rev. A}\ }\textbf {\bibinfo {volume} {46}},\ \bibinfo {pages} {380} (\bibinfo
  {year} {1992})}\BibitemShut {NoStop}%
\bibitem [{\citenamefont {Uzdin}\ and\ \citenamefont
  {Moiseyev}(2010)}]{Uzdi10}%
  \BibitemOpen
  \bibfield  {author} {\bibinfo {author} {\bibfnamefont {R.}~\bibnamefont
  {Uzdin}}\ and\ \bibinfo {author} {\bibfnamefont {N.}~\bibnamefont
  {Moiseyev}},\ }\href@noop {} {\bibfield  {journal} {\bibinfo  {journal}
  {Phys.~Rev.~A}\ }\textbf {\bibinfo {volume} {81}},\ \bibinfo {pages} {063405}
  (\bibinfo {year} {2010})}\BibitemShut {NoStop}%
\bibitem [{\citenamefont {Botheron}\ and\ \citenamefont {Pons}(2009)}]{Both09}%
  \BibitemOpen
  \bibfield  {author} {\bibinfo {author} {\bibfnamefont {P.}~\bibnamefont
  {Botheron}}\ and\ \bibinfo {author} {\bibfnamefont {B.}~\bibnamefont
  {Pons}},\ }\href@noop {} {\bibfield  {journal} {\bibinfo  {journal}
  {Phys.~Rev.~A}\ }\textbf {\bibinfo {volume} {80}},\ \bibinfo {pages} {023402}
  (\bibinfo {year} {2009})}\BibitemShut {NoStop}%
\bibitem [{\citenamefont {Yakovlev}\ and\ \citenamefont
  {Scrinzi}(2003)}]{Yako03}%
  \BibitemOpen
  \bibfield  {author} {\bibinfo {author} {\bibfnamefont {V.~S.}\ \bibnamefont
  {Yakovlev}}\ and\ \bibinfo {author} {\bibfnamefont {A.}~\bibnamefont
  {Scrinzi}},\ }\href@noop {} {\bibfield  {journal} {\bibinfo  {journal}
  {Phys.~Rev.~Lett.}\ }\textbf {\bibinfo {volume} {91}},\ \bibinfo {pages}
  {153901} (\bibinfo {year} {2003})}\BibitemShut {NoStop}%
\bibitem [{\citenamefont {Pukhov}\ \emph {et~al.}(2003)\citenamefont {Pukhov},
  \citenamefont {Gordienko},\ and\ \citenamefont {Baeva}}]{Pukh03}%
  \BibitemOpen
  \bibfield  {author} {\bibinfo {author} {\bibfnamefont {A.}~\bibnamefont
  {Pukhov}}, \bibinfo {author} {\bibfnamefont {S.}~\bibnamefont {Gordienko}}, \
  and\ \bibinfo {author} {\bibfnamefont {T.}~\bibnamefont {Baeva}},\
  }\href@noop {} {\bibfield  {journal} {\bibinfo  {journal} {Phys.~Rev.~Lett.}\
  }\textbf {\bibinfo {volume} {91}},\ \bibinfo {pages} {173002} (\bibinfo
  {year} {2003})}\BibitemShut {NoStop}%
\bibitem [{\citenamefont {Zagoya}\ \emph {et~al.}(2012)\citenamefont {Zagoya},
  \citenamefont {Goletz}, \citenamefont {Grossmann},\ and\ \citenamefont
  {Rost}}]{Zago12}%
  \BibitemOpen
  \bibfield  {author} {\bibinfo {author} {\bibfnamefont {C.}~\bibnamefont
  {Zagoya}}, \bibinfo {author} {\bibfnamefont {C.-M.}\ \bibnamefont {Goletz}},
  \bibinfo {author} {\bibfnamefont {F.}~\bibnamefont {Grossmann}}, \ and\
  \bibinfo {author} {\bibfnamefont {J.-M.}\ \bibnamefont {Rost}},\ }\href
  {http://stacks.iop.org/1367-2630/14/i=9/a=093050} {\bibfield  {journal}
  {\bibinfo  {journal} {New J. Phys.}\ }\textbf {\bibinfo {volume} {14}},\
  \bibinfo {pages} {093050} (\bibinfo {year} {2012})}\BibitemShut {NoStop}%
\bibitem [{\citenamefont {Kamor}\ \emph {et~al.}(2014)\citenamefont {Kamor},
  \citenamefont {Chandre}, \citenamefont {Uzer},\ and\ \citenamefont
  {Mauger}}]{Kamo14}%
  \BibitemOpen
  \bibfield  {author} {\bibinfo {author} {\bibfnamefont {A.}~\bibnamefont
  {Kamor}}, \bibinfo {author} {\bibfnamefont {C.}~\bibnamefont {Chandre}},
  \bibinfo {author} {\bibfnamefont {T.}~\bibnamefont {Uzer}}, \ and\ \bibinfo
  {author} {\bibfnamefont {F.}~\bibnamefont {Mauger}},\ }\href {\doibase
  10.1103/PhysRevLett.112.133003} {\bibfield  {journal} {\bibinfo  {journal}
  {Phys.~Rev.~Lett.}\ }\textbf {\bibinfo {volume} {112}},\ \bibinfo {pages}
  {133003} (\bibinfo {year} {2014})}\BibitemShut {NoStop}%
\bibitem [{\citenamefont {Valentin}\ \emph {et~al.}(2004)\citenamefont
  {Valentin}, \citenamefont {Kazamias}, \citenamefont {Douillet}, \citenamefont
  {Grillon}, \citenamefont {Lefrou}, \citenamefont {Aug{\'e}}, \citenamefont
  {Lewenstein}, \citenamefont {Wyart}, \citenamefont {Sebban},\ and\
  \citenamefont {Balcou}}]{Vale04}%
  \BibitemOpen
  \bibfield  {author} {\bibinfo {author} {\bibfnamefont {C.}~\bibnamefont
  {Valentin}}, \bibinfo {author} {\bibfnamefont {S.}~\bibnamefont {Kazamias}},
  \bibinfo {author} {\bibfnamefont {D.}~\bibnamefont {Douillet}}, \bibinfo
  {author} {\bibfnamefont {G.}~\bibnamefont {Grillon}}, \bibinfo {author}
  {\bibfnamefont {T.}~\bibnamefont {Lefrou}}, \bibinfo {author} {\bibfnamefont
  {F.}~\bibnamefont {Aug{\'e}}}, \bibinfo {author} {\bibfnamefont
  {M.}~\bibnamefont {Lewenstein}}, \bibinfo {author} {\bibfnamefont
  {J.}~\bibnamefont {Wyart}}, \bibinfo {author} {\bibfnamefont
  {S.}~\bibnamefont {Sebban}}, \ and\ \bibinfo {author} {\bibfnamefont
  {P.}~\bibnamefont {Balcou}},\ }\href@noop {} {\bibfield  {journal} {\bibinfo
  {journal} {J.~Phys.~B}\ }\textbf {\bibinfo {volume} {37}},\ \bibinfo {pages}
  {2661} (\bibinfo {year} {2004})}\BibitemShut {NoStop}%
\bibitem [{\citenamefont {Higuet}\ \emph {et~al.}(2011)\citenamefont {Higuet},
  \citenamefont {Ruf}, \citenamefont {Thir\'e}, \citenamefont {Cireasa},
  \citenamefont {Constant}, \citenamefont {Cormier}, \citenamefont {Descamps},
  \citenamefont {M\'evel}, \citenamefont {Petit}, \citenamefont {Pons},
  \citenamefont {Mairesse},\ and\ \citenamefont {Fabre}}]{Higu11}%
  \BibitemOpen
  \bibfield  {author} {\bibinfo {author} {\bibfnamefont {J.}~\bibnamefont
  {Higuet}}, \bibinfo {author} {\bibfnamefont {H.}~\bibnamefont {Ruf}},
  \bibinfo {author} {\bibfnamefont {N.}~\bibnamefont {Thir\'e}}, \bibinfo
  {author} {\bibfnamefont {R.}~\bibnamefont {Cireasa}}, \bibinfo {author}
  {\bibfnamefont {E.}~\bibnamefont {Constant}}, \bibinfo {author}
  {\bibfnamefont {E.}~\bibnamefont {Cormier}}, \bibinfo {author} {\bibfnamefont
  {D.}~\bibnamefont {Descamps}}, \bibinfo {author} {\bibfnamefont
  {E.}~\bibnamefont {M\'evel}}, \bibinfo {author} {\bibfnamefont
  {S.}~\bibnamefont {Petit}}, \bibinfo {author} {\bibfnamefont
  {B.}~\bibnamefont {Pons}}, \bibinfo {author} {\bibfnamefont {Y.}~\bibnamefont
  {Mairesse}}, \ and\ \bibinfo {author} {\bibfnamefont {B.}~\bibnamefont
  {Fabre}},\ }\href {\doibase 10.1103/PhysRevA.83.053401} {\bibfield  {journal}
  {\bibinfo  {journal} {Phys. Rev. A}\ }\textbf {\bibinfo {volume} {83}},\
  \bibinfo {pages} {053401} (\bibinfo {year} {2011})}\BibitemShut {NoStop}%
\bibitem [{\citenamefont {Mart{\'\i}nez}\ \emph
  {et~al.}(2015)\citenamefont {Mart{\'\i}nez}, \citenamefont
  {Babushkin}, \citenamefont {Berg{\'e}}, \citenamefont {Skupin}, \citenamefont
  {Cabrera-Granado}, \citenamefont {K{\"o}hler}, \citenamefont {Morgner},
  \citenamefont {Husakou},\ and\ \citenamefont {Herrmann}}]{Mart15}%
  \BibitemOpen
  \bibfield  {author} {\bibinfo {author} {\bibfnamefont {P.~G.}\ \bibnamefont
  {Mart{\'\i}nez}}, \bibinfo {author} {\bibfnamefont
  {I.}~\bibnamefont {Babushkin}}, \bibinfo {author} {\bibfnamefont
  {L.}~\bibnamefont {Berg{\'e}}}, \bibinfo {author} {\bibfnamefont
  {S.}~\bibnamefont {Skupin}}, \bibinfo {author} {\bibfnamefont
  {E.}~\bibnamefont {Cabrera-Granado}}, \bibinfo {author} {\bibfnamefont
  {C.}~\bibnamefont {K{\"o}hler}}, \bibinfo {author} {\bibfnamefont
  {U.}~\bibnamefont {Morgner}}, \bibinfo {author} {\bibfnamefont
  {A.}~\bibnamefont {Husakou}}, \ and\ \bibinfo {author} {\bibfnamefont
  {J.}~\bibnamefont {Herrmann}},\ }\href@noop {} {\bibfield  {journal}
  {\bibinfo  {journal} {Phys.~Rev.~Lett.}\ }\textbf {\bibinfo {volume} {114}},\
  \bibinfo {pages} {183901} (\bibinfo {year} {2015})}\BibitemShut {NoStop}%
\bibitem [{\citenamefont {Schuh}\ \emph {et~al.}(2017)\citenamefont {Schuh},
  \citenamefont {Kolesik}, \citenamefont {Wright}, \citenamefont {Moloney},\
  and\ \citenamefont {Koch}}]{Schu17}%
  \BibitemOpen
  \bibfield  {author} {\bibinfo {author} {\bibfnamefont {K.}~\bibnamefont
  {Schuh}}, \bibinfo {author} {\bibfnamefont {M.}~\bibnamefont {Kolesik}},
  \bibinfo {author} {\bibfnamefont {E.~M.}\ \bibnamefont {Wright}}, \bibinfo
  {author} {\bibfnamefont {J.~V.}\ \bibnamefont {Moloney}}, \ and\ \bibinfo
  {author} {\bibfnamefont {S.~W.}\ \bibnamefont {Koch}},\ }\href@noop {}
  {\bibfield  {journal} {\bibinfo  {journal} {Phys.~Rev.~Lett.}\ }\textbf
  {\bibinfo {volume} {118}},\ \bibinfo {pages} {063901} (\bibinfo {year}
  {2017})}\BibitemShut {NoStop}%
\bibitem [{\citenamefont {Evstatiev}\ and\ \citenamefont
  {Shadwick}(2013)}]{Evst13}%
  \BibitemOpen
  \bibfield  {author} {\bibinfo {author} {\bibfnamefont {E.~G.}\ \bibnamefont
  {Evstatiev}}\ and\ \bibinfo {author} {\bibfnamefont {B.~A.}\ \bibnamefont
  {Shadwick}},\ }\href@noop {} {\bibfield  {journal} {\bibinfo  {journal} {J.
  Comput. Phys.}\ }\textbf {\bibinfo {volume} {245}},\ \bibinfo {pages} {376}
  (\bibinfo {year} {2013})}\BibitemShut {NoStop}%
\bibitem [{\citenamefont {Ruth}(1983)}]{Ruth83}%
  \BibitemOpen
  \bibfield  {author} {\bibinfo {author} {\bibfnamefont {R.}~\bibnamefont
  {Ruth}},\ }\href@noop {} {\bibfield  {journal} {\bibinfo  {journal} {IEEE
  Trans. Nucl. Sci.}\ }\textbf {\bibinfo {volume} {30}},\ \bibinfo {pages}
  {2669} (\bibinfo {year} {1983})}\BibitemShut {NoStop}%
\end{thebibliography}
\end{document}


\title{Coherent buildup of high harmonic radiation: the classical perspective\\
Supplemental Material}

\date{\today}

\author{S.A.\ Berman$^{1,2}$, J. Dubois$^2$, C.\ Chandre$^2$, M.\ Perin$^1$, T.\ Uzer$^1$}
\affiliation{$^1$School of Physics, Georgia Institute of Technology, Atlanta, Georgia 30332-0430, USA}
\affiliation{$^2$Aix Marseille Univ, CNRS, Centrale Marseille, I2M, Marseille, France}

\maketitle
\section{Numerical solution of the model equations}
In this section, we describe the numerical methods we employ to solve our model equations.
Equation (1) in the paper is a partial differential equation for the electric field $\mathcal{E}(\tau,z)$, where $z$ is the evolution parameter.
The right-hand side of the equation contains the ensemble-averaged dipole velocity $\langle v(\tau,z) \rangle$, which represents the response of the atoms to the field at each $z$.
\begin{figure}
\includegraphics[width=0.4\textwidth]{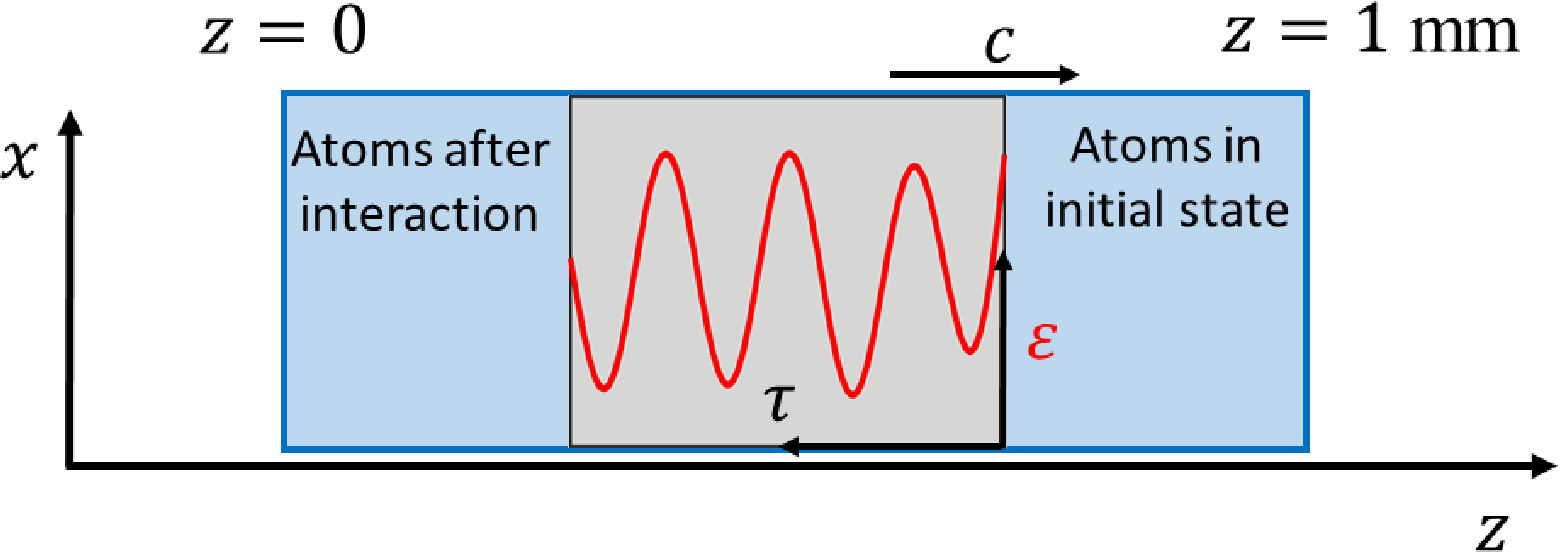}
\caption{Schematic of the model, Eq.~(1) in the paper.}\label{fig:schematic}
\end{figure}
As illustrated in Fig.~\ref{fig:schematic}, $\langle v(\tau,z) \rangle$ must be obtained for a given $z$, whether classically or quantum-mechanically, and this allows the electric field to be advanced one step $\Delta z$ in space to $z+\Delta z$.
This space-stepping is performed over the entire length of the gas, from $z=0$ to $z=1\,\,\mathrm{mm}$, using a third-order implicit multistep method.

To obtain $\langle v(\tau,z) \rangle$ classically, we must first solve Eq.~(2) in the paper, which is the Liouville equation
\begin{equation}\label{eq:liouville}
\frac{\partial f}{\partial \tau} = - v \frac{\partial f}{\partial x}+ \left( \frac{\partial V}{\partial x} + \mathcal{E}(\tau,z) \right) \frac{\partial f}{\partial v}.
\end{equation}
This may also be written as
\begin{equation}
\frac{\partial f}{\partial \tau} = - \{f,H\}
\end{equation}
where $H(x,v,\tau;z)$ is Hamiltonian (3) in the paper and $\{\cdot,\cdot\}$ is the Poisson bracket for canonically conjugate variables $(x,v)$,
\begin{equation}
\{F,G\} = \frac{\partial F}{\partial x}\frac{\partial G}{\partial v} - \frac{\partial F}{\partial v}\frac{\partial G}{\partial x}.
\end{equation}
We solve Eq.~\eqref{eq:liouville} using the scheme developed in Ref.~\cite{Evst13}: we express $f$ as 
\begin{equation*}
f(x,v,\tau;z) = \sum_{\alpha=1}^N w_\alpha \delta(x - x_\alpha(\tau;z)) \delta(v - v_\alpha(\tau;z)),
\end{equation*}
where $\delta$ is the Dirac delta function and the $w_\alpha$ are constant weighting factors satisfying $\sum_\alpha w_\alpha = 1$.
The $(x_\alpha(\tau;z),v_\alpha(\tau;z))$ are classical trajectories obtained by numerically solving the equations of motion associated with $H(x,v,\tau;z)$, using a third-order explicit symplectic scheme \cite{Ruth83}.
Having thus solved Eq.~\eqref{eq:liouville} for a given $z$, the ensemble-averaged dipole velocity is obtained as
\begin{equation}
\langle v(\tau,z) \rangle = \int v f(x,v,\tau;z) \mathrm{d}x\mathrm{d}v.
\end{equation}
This approach is very similar to the classical trajectory Monte Carlo approach \cite{Sand99,Band92,Both09,Higu11}.
The main difference is that the initial conditions $(x_\alpha(0;z),v_\alpha(0;z))$ are chosen on a uniform grid, improving convergence with increasing $N$ compared to the Monte Carlo case \cite{Uzdi10}, and they are weighted such that the weights $w_\alpha \propto f_0(x_\alpha(0;z),v_\alpha(0;z))$ correspond to a given distribution function $f_0$ representing the initial state of the atom.
For the simulations presented in the paper, we used $N \sim 4 \times 10^6$ trajectories.

To obtain $\langle v(\tau,z) \rangle$ quantum-mechanically, the following time-dependent Schr\"{o}dinger equation must first be solved for the wavefunction $\psi(x,\tau;z)$:
\begin{equation}\label{eq:schrod}
i \frac{\partial \psi}{\partial \tau} = - \frac{1}{2}\frac{\partial^2 \psi}{\partial x^2} + \big(V(x) + \mathcal{E}(\tau,z) x \big) \psi.
\end{equation} 
We use a split-operator method to numerically solve Eq.~\eqref{eq:schrod}.
The dipole velocity is then obtained from $\psi$ as the integral of the dipole acceleration, 
\begin{align}
\langle v(\tau,z) \rangle & = \langle v(0,z) \rangle \nonumber \\
- & \int_0^\tau \left\{ \mathcal{E}(\tau',z) +  \int \frac{\partial V}{ \partial x} |\psi(x,\tau';z)|^2\,\mathrm{d}x \right\}\,\mathrm{d} \tau'.
\end{align}

\section{Computation of the observables}
To perform the time-frequency analysis of an observable $O(\tau;z)$, we compute
\begin{equation}
\tilde{O}(\tau,\omega;z) = \left| \int W_T(\tau' - \tau) O(\tau';z) \exp(-i \omega \tau') \mathrm{d} \tau' \right|^2.
\end{equation}
We take the window function to be $W_T(\tau)=\cos^4(\pi\tau/T)$ for the range $-T/2 < \tau < T/2$ and zero elsewhere.
This is done in Fig.~2 in the paper for the dipole acceleration $-\langle \partial V / \partial x (\tau,z)\rangle$ computed from the quantum model, with $T=0.35\,\,\mathrm{l.c.}$
To obtain the time-dependent kinetic energy distributions of recolliding electrons from the classical model, we compute
\begin{align}
R(\kappa,\tau;z) = & \int f(x,v,\tau;z) \Theta(x_c-|x|) \nonumber \\
\times & \Theta(\kappa_c U_p - |v^2/2-\kappa U_p|)\,\mathrm{d}x\mathrm{d}v,
\end{align}
where $\Theta$ is the Heaviside step function and $x_c = 5\,\,\mathrm{a.u.}$ and $\kappa_c=0.014$ are thresholds.

Because our distribution function $f(x,v,\tau;z)$ is expressed as a sum over Dirac delta functions, it must be coarse-grained in order to make the plots in Fig.~3 of the paper.
Thus, what is actually plotted in this figure is
\begin{align}
\tilde{f}(x_g,v_g,& \tau;z) = (\Delta x \Delta v)^{-1}  \int f(x,v,\tau;z) \nonumber \\
\times & \Theta(\Delta x/2 - |x - x_g|) \Theta (\Delta v / 2 - |v - v_g| ) \mathrm{d}x\mathrm{d}v
\end{align}
for $x_g$ and $v_g$ on a fine grid of spacing $\Delta x$ and $\Delta v$ in the $x$ and $v$ directions, respectively.